\pdfoutput=1

\documentclass[11pt]{article}

\usepackage[final]{acl}  

\usepackage{times}
\usepackage{latexsym}

\usepackage[T1]{fontenc}

\usepackage[utf8]{inputenc}

\usepackage{microtype}

\usepackage{inconsolata}

\usepackage{mdframed}
\usepackage{graphicx}
\usepackage{float}
\usepackage{hyperref}
\usepackage{url}

\usepackage{amssymb}
\usepackage{amsmath}
\usepackage{epsfig,subfigure,caption}
\usepackage{algpseudocode}
\usepackage[normalem]{ulem}
\usepackage[linesnumbered,algoruled,boxed,noend]{algorithm2e}
\usepackage{listings} 

\usepackage{xcolor}
\usepackage{color, colortbl}
\usepackage{multirow,booktabs, hhline}
\usepackage{bm}
\usepackage[linesnumbered,algoruled,boxed,noend]{algorithm2e}
\usepackage{wrapfig}
\usepackage{verbatimbox}
\usepackage{kantlipsum}
\usepackage{fancyvrb}
\usepackage[htt]{hyphenat}
\usepackage{pifont}
\newcommand{\cmark}{\ding{51}}%
\newcommand{\xmark}{\ding{55}}%

\usepackage{tcolorbox}
\usepackage{enumitem}

\tcbset{
    myboxstyle/.style={
        colback=yellow!10!white, 
        colframe=yellow!50!black, 
        fonttitle=\bfseries,
        coltitle=black,
        boxrule=0.75mm,
        width=\textwidth,
        sharp corners,
        leftrule=1mm,
        left=5pt,
        right=5pt,
        top=5pt,
        bottom=5pt
    }
}
\newcommand{\secref}[1]{\S\ref{#1}}

\title{STAR: A Simple Training-free Approach for \\ Recommendations using Large Language Models}

\author{
    Dong-Ho Lee\textsuperscript{1}\thanks{~~Authors contributed equally.}\thanks{~~Work done during internship at Google DeepMind.},~
    Adam Kraft\textsuperscript{2}$^*$,~
    Long Jin\textsuperscript{3},~
    Nikhil Mehta\textsuperscript{2},~\\
    \textbf{Taibai Xu\textsuperscript{3}}~,
    \textbf{Lichan Hong\textsuperscript{2}},~
    \textbf{Ed H. Chi\textsuperscript{2}},~
    \textbf{Xinyang Yi\textsuperscript{2}},~
    \\
    \textsuperscript{1}Information Sciences Institute, University of Southern California \\
    \textsuperscript{2}Google DeepMind~
    \textsuperscript{3}Google \\
    {\small \texttt{\{dongho.lee\}@usc.edu}, \texttt{\{adamkraft,longjin,nikhilmehta,taibaixu,lichan,edchi,xinyang\}@google.com}}\\
}

\begin{document}

\maketitle
\begin{abstract}
Recent progress in large language models (LLMs) offers promising new approaches for recommendation system tasks.
While the current state-of-the-art methods rely on fine-tuning LLMs to achieve optimal results, this process is costly and introduces significant engineering complexities. 
Conversely, methods that directly use LLMs without additional fine-tuning result in a large drop in recommendation quality, often due to the inability to capture collaborative information.
In this paper, we propose a \textbf{S}imple \textbf{T}raining-free \textbf{A}pproach for \textbf{R}ecommendation (\textbf{STAR}), a framework that utilizes LLMs and can be applied to various recommendation tasks without the need for fine-tuning, while maintaining high quality recommendation performance.
Our approach involves a retrieval stage that uses semantic embeddings from LLMs combined with collaborative user information to retrieve candidate items.
We then apply an LLM for pairwise ranking to enhance next-item prediction.
Experimental results on the Amazon Review dataset show competitive performance for next item prediction, even with our retrieval stage alone. Our full method achieves Hits@10 performance of +23.8\% on \textit{Beauty}, +37.5\% on \textit{Toys \& Games}, and -1.8\% on \textit{Sports \& Outdoors} relative to the best supervised models.
This framework offers an effective alternative to traditional supervised models, highlighting the potential of LLMs in recommendation systems without extensive training or custom architectures.
\end{abstract}

\section{Introduction}
\begin{figure}[t!]
    \centering
    \begin{minipage}{\columnwidth}
    \centering
    \includegraphics[width=0.9\columnwidth]{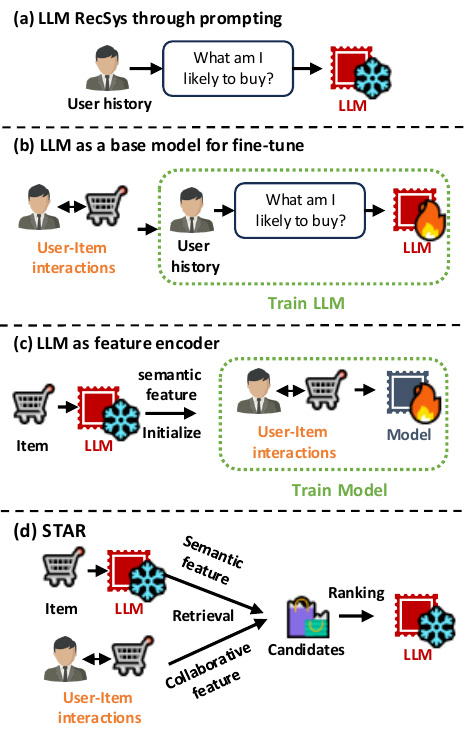}
    \vspace{-0.0cm}
\end{minipage}
\caption{\textbf{A Motivating Example.} LLMs can be utilized in RecSys through (a) prompting, (b) fine-tuning on user-item interactions, and (c) using LLMs as feature encoders for training subsequent models. However, (a) cannot leverage collaborative knowledge, while (b) and (c) require extensive training and large-scale interaction data. Our framework \texttt{STAR} integrates collaborative knowledge into LLMs without additional training.}
    \label{fig:motivation}
\end{figure}
Recent advances in large language models (LLMs) present new opportunities for addressing recommendation tasks~\citep{brown2020language, team2023gemini, lin2023can, zhao2023recommender, li2023large, chen2024large, tsai-etal-2024-leveraging, wu2024survey}.
Current approaches primarily leverage LLMs in three ways:
(1) advanced \textit{prompting}~\cite{wang2023zero, wang2023drdt, hou2024large, wang-etal-2024-recmind, xu2024prompting, zhao2024let, liang2024taxonomy};
(2) as base models for direct \textit{fine-tuning}~\cite{geng2022recommendation, zhang2023recommendation, bao2023tallrec,  xu2024openp5, tan2024idgenrec, kim2024large}; and
(3) as feature encoders for \textit{fine-tuning} subsequent generative models~\citep{hou2023learning, singh2023better, rajput2024recommender, zheng2024adapting} or sequential models~\cite{sun2019bert4rec, yuan2023go, hu2024enhancing}.
While \textit{prompting} LLMs for scoring and ranking utilizes their reasoning abilities, these models often largely underperform compared to fine-tuned approaches due to the absence of \textbf{collaborative} knowledge derived from user-item interactions.
Conversely, using LLMs for \textit{fine-tuning}, whether as base models or feature encoders, enhances model performance by leveraging their strong \textbf{semantic} understanding. 
However, this requires extensive training and large-scale interaction data.

The main objective of this work is to integrate the semantic capabilities of LLMs with collaborative knowledge from user-item interaction data, all without requiring additional training.
To achieve this, we present a \textbf{S}imple \textbf{T}raining-free \textbf{A}pproach for \textbf{R}ecommendation (\textbf{STAR}) framework using LLMs.
The STAR framework involves two stages: \textbf{Retrieval} and \textbf{Ranking}. The \textbf{Retrieval} stage scores new items using a combination of \textit{semantic similarity} and \textit{collaborative commonality} to the items in a user's history. Here, we utilize LLM-based embeddings to determine semantic similarity. Additionally, a \textit{temporal} factor gives priority to user's recent interactions, and a \textit{rating} factor aligns with user preferences to rank items within a specific set (\secref{ssec:retrieval}).
The \textbf{Ranking} stage leverages the reasoning capabilities of LLMs to adjust the rankings of the initially retrieved candidates.
Specifically, we assess various LLM-based ranking approaches, including point-wise, pair-wise, and list-wise methods, while also determining the key information needed for the LLM to better understand user preferences and make accurate predictions (\secref{ssec:reranking}).
Our experimental evaluation shows competitive performance across a diverse range of recommendation datasets, all without the need for supervised training or the development of custom-designed architectures.

We present extensive experimental results
on the Amazon Review dataset~\citep{mcauley2015image, he2016ups}.
Our findings are as follow:

\begin{enumerate}
  \item Our retrieval pipeline itself, comprised of both semantic relationship and collaborative information, achieves Hits@10 performance of +17.3\% on \textit{Beauty}, +26.2\% on \textit{Toys \& Games}, and -5.5\% on \textit{Sports \& Outdoors} relative to the best supervised models.
  \item We show that pair-wise ranking further improves upon our retrieval performance, while point-wise and list-wise methods struggle to achieve similar improvements.
  \item We illustrate that collaborative information is a critical component that adds additional benefits to the semantic information throughout our system, in both the retrieval and ranking stages.
\end{enumerate}


These findings show that it is possible to build a recommendation system utilizing LLMs without additional fine-tuning that can significantly close the quality gap of fully fine-tuned systems, and in many cases even achieve higher quality.
\section{Related Works}

Recent studies have explored the role of LLMs in recommendation systems through three primary approaches: (1) using prompting for scoring and ranking, (2) fine-tuning as a base model, and (3) serving as a feature encoder.

\paragraph{LLM prompting for scoring and ranking.}
LLMs can generate recommendations by understanding user preferences or past interactions expressed in natural language.

This is typically achieved through generative selection prompting, where the model ranks and selects top items from a given candidate set \citep{wang2023zero, wang2023drdt, hou2024large, wang-etal-2024-recmind, xu2024prompting, zhao2024let, liang2024taxonomy}.
Another line of work applies ranking prompting \citep{dai2023uncovering}, inspired by LLM-based ranking in information retrieval \citep{zhu2023large, wang2024large}, using point-wise \citep{liang2022holistic, zhuang2023beyond}, pair-wise \citep{qin-etal-2024-large}, or list-wise \citep{sun2023chatgpt, qin-etal-2024-large} approaches.
However, LLM prompting alone, without a retrieval stage to pre-select candidate items for scoring and ranking, is less effective than fine-tuned models due to lack of collaborative knowledge derived from user-item interaction data.
As a result, many approaches use a fine-tuned model for candidate retrieval and an LLM for ranking. 
However, in this setup, overall performance is primarily determined by the retrieval quality rather than the LLM itself.

\paragraph{LLM as base model for fine-tuning.}
To integrate collaborative knowledge with the semantic understanding of LLMs, recent studies have explored fine-tuning using user-item interaction data.
While this improves recommendation performance, it requires extensive training and large-scale interaction datasets \citep{geng2022recommendation, zhang2023recommendation, bao2023tallrec, xu2024openp5, tan2024idgenrec, kim2024large}.

\paragraph{LLM as feature encoder.}
LLMs can also be used as text encoders to capture richer semantic information from item metadata and user profiles~\cite{reimers2019sentence, cer-etal-2018-universal, ni2021sentence, lee2024gecko}.
To further optimize these representations, researchers have explored several approaches:
(1) mapping continuous LLM embeddings into discrete tokens using vector quantization and training a subsequent generative model~\citep{hou2023learning, singh2023better, rajput2024recommender, zheng2024adapting};
(2) training sequential models by initializing the embedding layer with LLM embeddings~\citep{sun2019bert4rec, yuan2023go, hu2024enhancing}; and
(3) training models to directly compute the relevance between item and user embeddings (\textit{i.e.,} embeddings of user selected items)~\citep{Ding2022, hou2022towards, gong2023unified, li2023text, liu2024once, li2024enhancing, ren2024representation, sheng2024language}.

In this work, we utilize LLMs as both feature encoders and ranking functions by integrating collaborative knowledge from user-item interaction data.
Our findings show that LLM embeddings can serve as effective item representations, achieving strong results in sequential recommendation tasks without extensive optimization.
This aligns with \citep{harte2023leveraging} but differs in our use of novel scoring rules that incorporate both collaborative and temporal information.


\section{STAR: Simple Training-Free RecSys}

\begin{figure*}[h]
  \begin{center}
    \includegraphics[width=\linewidth]{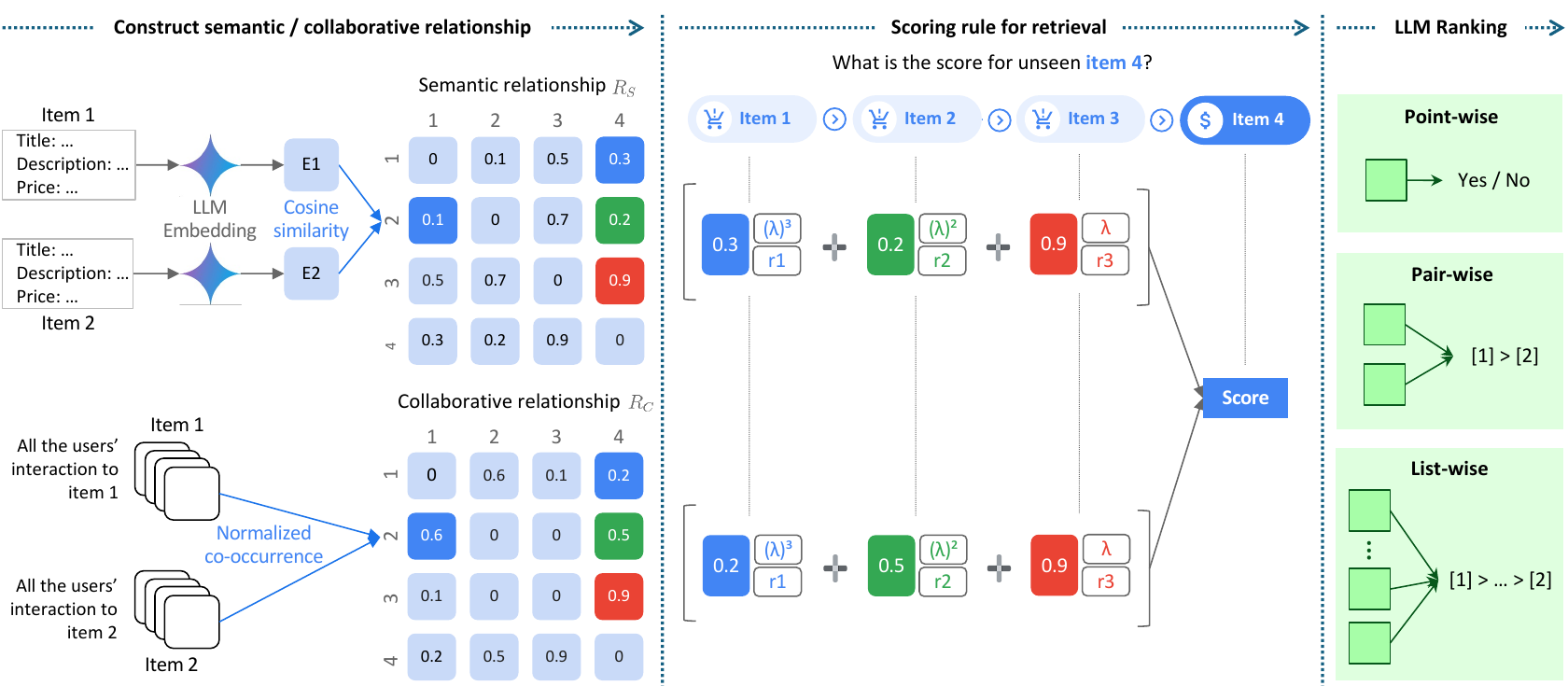}
  \end{center}
\vspace{-0.4cm}
  \caption{\textbf{STAR Framework overview.} We use the semantic relationship scores in $R_{\text{S}}$ and the collaborative relationship scores in $R_{\text{C}}$ to score the items in the user history compared to new items to recommend. The final score for one new item is a weighted average from the semantic relationship and collaborative relationship scores, with additional weights from the user's ratings $r$ and a temporal decay $\lambda<1$ which prioritize recent interactions. The top scoring retrieved items are sent to the LLM Ranking, where we can use point-wise, pair-wise, or list-wise ranking approaches to further improve upon the scoring of recommended items.
  }
  \label{fig:framework-overview}
\end{figure*}

This section initially outlines the problem formulation (\secref{ssec:formulation}). 
Subsequently, we detail the proposed retrieval (\secref{ssec:retrieval}) and ranking pipelines (\secref{ssec:reranking}).

\subsection{Sequential Recommendation}
\label{ssec:formulation}
The sequential recommendation task predicts the next item a user will interact with based on their interaction history.
For a user $u \in U$, where $U$ is the set of all users, the interaction history is represented as a sequence of items $S_u = \{s_1, s_2, \dots, s_n\}$, with each item $s_i \in I$ belonging to the set of all items $I$.
Each user history item $s_i$ is associated with a rating $r_i \in \{1, 2, 3, 4, 5\}$ given by the user $u$.
The goal is to predict the next item $s_{n+1} \in I$ that the user is most likely to interact with.

\subsection{Retrieval Pipeline}
\label{ssec:retrieval}
The retrieval pipeline aims to assign a score to an unseen item $x \in I$ given the sequence $S_u$.
To achieve this, we build two scoring components: one that focuses on the \textbf{semantic} relationship between items and another that focuses on the \textbf{collaborative} relationship.


\paragraph{Semantic relationship.}
Understanding how similar a candidate item is to the items in a user's interaction history $s_i \in S_u$ is key to accurately gauging how well candidate items align with user preferences. Here we leverage LLM embedding models, where we pass in text prompts representing items and collect embedding vectors of dimension $d_e$. We construct a prompt based on the item information and metadata, which can include fields like \textit{title}, \textit{description}, \textit{category}, \textit{brand}, \textit{sales ranking}, \textit{price}, etc. (See Appendix~\ref{app:item-prompt} for the full prompt details).
We collect embeddings for each item $i \in I$, resulting in $E \in \mathbb{R}^{n \times d_e}$, where $n$ is number of total items in $I$.

The semantic relationship between two items ($i_a, i_b$) is then calculated using the cosine similarity between their embeddings $E_{i_a}, E_{i_b} \in E$. 
This measure provides a numerical representation of how closely related the items are in semantic space. 
For our experiments, we precompute the entire semantic relationship matrix $R_{\text{S}} \in \mathbb{R}^{n \times n}$. 
For many domains, this is a practical solution. 
However, if $|I|$ is very large, Approximate Nearest Neighbor methods~\citep{guo2020accelerating, sun2024soar} are efficient approaches to maintain quality and reduce computation.

\paragraph{Collaborative relationship.}
Semantic similarity between a candidate item and items in a user's interaction history is a helpful cue for assessing the similarity of items based on the item information.
However, this alone does not fully capture the engagement interactions of items by multiple users.
To better understand the collaborative relationship, we consider how frequently different combinations of items are interacted with by users. 
These shared interaction patterns can provide strong indicators of how likely the candidate item is to resonate with a broader audience with similar preferences.
For each item $i \in I$, we derive an interaction array that represents user interactions, forming a set of sparse user-item interaction arrays $C \in \mathbb{R}^{n \times m}$, where $m$ is number of users in $U$.
The collaborative relationship between two items ($i_a, i_b$) is then computed by using the cosine similarity between their sparse arrays $C_{i_a}, C_{i_b} \in C$, capturing the normalized co-occurrence of the items.
To streamline the process, we pre-compute and store these values in a collaborative relationship matrix $R_{\text{C}} = \frac{C \cdot  C^\top}{\|C\| \|C^\top\|} \in \mathbb{R}^{n \times n}$, which is typically very sparse.

\paragraph{Scoring rules.}
The score for an unseen item $x \in I$ is calculated by averaging both the semantic and collaborative relationships between items in $S_u = \{s_1, s_2, \dots, s_n\}$ as follows:
\begin{equation} \label{eq:1}
    \texttt{score}(x) = \frac{1}{n} \sum_{j=1}^{n} r_j  \lambda^{t_j}  \left[ a  R_{\text{S}}^{xj} + (1-a)  R_{\text{C}}^{xj} \right]
\end{equation}

where $R_{\text{S}}^{xj}$ and $R_{\text{C}}^{xj}$ represent the semantic and collaborative relationships between the unseen item $x$ and item $s_j \in S_u$, respectively.
In this equation, $r_j$ is the rating given by user $u$ to item $s_j$, and $\lambda^{t_j}$ is an exponential decay function applied to the temporal order ${t_j}$ of $s_j$ in the sequence $S_u$.
Here, ${t_j}$ is set to 1 for the most recent item in $S_u$ and increments by 1 up to $n$ for the oldest item.
The framework, illustrated in Figure \ref{fig:framework-overview}, outputs the top $k$ items in descending order based on their scores.

\subsection{Ranking Pipeline}
\label{ssec:reranking}
After retrieving the top $k$ items, denoted as $I_k$, from the initial retrieval process, a LLM is employed to further rank these items to enhance the overall next-item recommendation quality.
The items in $I_k$ are already ordered based on scores from the retrieval framework, which reflect \textit{semantic}, \textit{collaborative}, and \textit{temporal} information. 
We intentionally incorporate this initial order into the ranking process to enhance both efficiency and effectiveness.
This framework then leverages the capabilities of the LLM to better capture user preference, complex relationships and contextual relevance among the items.

\subsubsection{Rank schema}
We present three main strategies for ranking:
(1) \textbf{Point-wise} evaluates each item $x \in I_k$ independently, based on the user sequence $S_u$, to determine how likely it is that user $u$ will interact with item $x$. 
If two items receive the same score, their rank follows the initial order from $I_k$;
(2) \textbf{Pair-wise} evaluates the preference between two items $x_i, x_j \in I_k$ based on the user sequence $S_u$. 
We adopt a sliding window approach, starting from the items with the lowest retrieval score at the bottom of the list~\citep{qin-etal-2024-large}. 
The LLM compares and swaps adjacent pairs, while iteratively stepping the comparison window one element at a time.
(3) \textbf{List-wise} evaluates the preference among multiple items $x_i, \dots, x_{i+w} \in I_k$ based on the user sequence $S_u$. This method also uses a sliding window approach, with a window size $w$ and a stride $d$ to move the window across the list, refining the ranking as it passes~\citep{sun2023chatgpt}. In this setup, \textbf{pair-wise} is a special case of \textbf{list-wise} with $w=2$ and $d=1$.
\begin{figure}[t]
  \begin{center}
    \includegraphics[width=\linewidth]{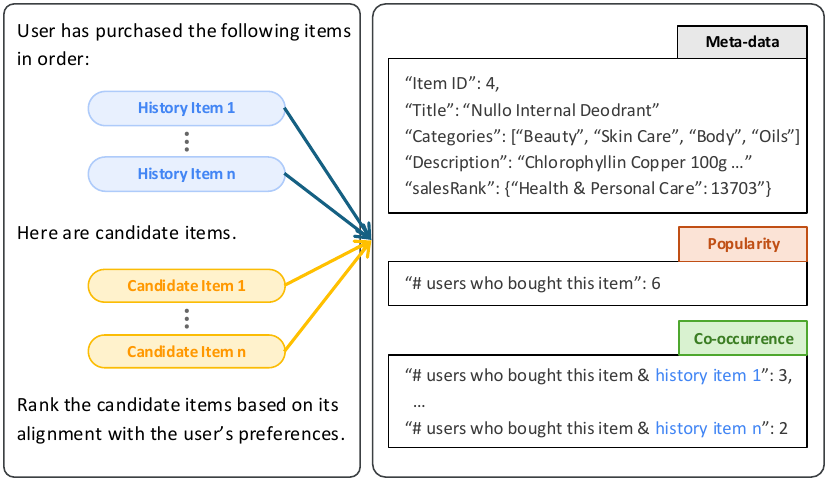}
  \end{center}
    \caption{\textbf{Prompt overview for the ranking pipeline.} The prompt includes history items, candidate items, and instructions for the ranking strategy. Each item is represented by metadata, along with additional details such as popularity and co-occurrence, formatted in JSON. Full prompt is available in Appendix~\ref{app:ranking-prompt}.}
\end{figure}

\subsubsection{Item information}
We represent the metadata (e.g., Item ID, title, category, etc.) for each item in the user sequence $s_j \in S_u$ and each candidate item to be ranked $x \in I_k$ as JSON format in the input prompt.
Additionally, we incorporate two more types of information that can help the reasoning capabilities of the LLM:
(1) \textbf{Popularity} is calculated as the number of users who have interacted with the item $x$, simply by counting the occurrences in the training data. 
This popularity value is then included in the prompt for both the items in the user sequence $s_j \in S_u$ and the candidate item to be ranked $x \in I_k$ as ``\textit{Number of users who interacted with this item: \#\#\#}'';
(2) \textbf{Co-occurrence} is calculated as the number of users who have interacted with both item $x$ and item $s_j \in S_u$.
The resulting value is then included for candidate items $x \in I_k$ as ``\textit{Number of users who interacted with both this item and item $s_j$: \#\#\#}''.
\section{Experimental Setup}

\paragraph{Datasets.} 
We evaluate the performance using public 2014 Amazon review datasets~\citep{mcauley2015image, he2016ups}. 
Specifically, we select the \textit{Beauty}, \textit{Toys and Games}, and \textit{Sports and Outdoors} categories, as these have been used in previous studies~\citep{geng2022recommendation, tan2024idgenrec} and provide data points for comparison (see Table~\ref{tab:dataset}). 
We follow the same data processing steps as in prior work, filtering out users and items with fewer than five interactions, maintaining consistent baseline settings.

\begin{table}[t]
    \begin{minipage}{0.48\textwidth}
    \centering
    \small
    \resizebox{\textwidth}{!}{
    	\begin{tabular}{lccccccc}
    	    \toprule
            Dataset & \# Users & \# Items & \# Interactions & Density \\
            \midrule
            Beauty & 22,363 & 12,101 & 198,502 & 0.0734\% \\
            Toys and Games & 19,412 & 11,924 & 167,597 & 0.0724\% \\
            Sports and Outdoors & 35,598 &  18,357 & 296,337 & 0.0453\% \\
            \bottomrule
        \end{tabular}
    }
    \end{minipage}
      \caption{\textbf{Dataset statistics}. Density is the percentage of actual user-item interactions out of all interactions.}
    \label{tab:dataset}
\end{table}

\paragraph{Dataset Construction and Evaluation Metrics.}
We follow conventional supervised models, where the last item, $s_n$, is reserved for testing and the second to last item, $s_{n-1}$, is used for validation. The remaining items are used for training. For the final predictions of $s_n$, all training and validation items are used as input. Although our method does not train model parameters, we only use the training data to calculate the collaborative user interaction values used for $R_C$ in retrieval and for \textbf{popularity} and \textbf{co-occurence} in ranking. We report Normalized Discounted Cumulative Gain (NDCG) and Hit Ratio (HR) at ranks 5 and 10.


\paragraph{Compared Methods.}
We compare our model with following supervised trained models:
(1) \textbf{KNN} is a user-based collaborative filtering method that finds the top 10 most similar users to a given user and averages their ratings to score a specific item;
(2) \textbf{Caser} uses convolution neural networks to model user interests~\citep{tang2018personalized};
(3) \textbf{HGN} uses hierarchical gating networks to capture both long and short-term user behaviors~\citep{ma2019hierarchical};
(4) \textbf{GRU4Rec} employs GRU to model user action sequences~\citep{hidasi2015session};
(5) \textbf{FDSA} uses a self-attentive model to learn feature transition patterns~\citep{zhang2019feature};
(6) \textbf{SASRec} uses a self-attention mechanism to capture item correlations within a user's action sequence~\citep{kang2018self};
(7) \textbf{BERT4Rec} applies a masked language modeling (MLM) objective for bi-directional sequential recommendation~\citep{sun2019bert4rec};
(8) \textbf{$\bf{S^{3}}$-Rec} extends beyond the MLM objective by pre-training with four self-supervised objectives to learn better item representations~\citep{zhou2020s3}.
(9) \textbf{P5}  fine-tunes a pre-trained LM for use in multi-task recommendation systems by generating tokens based on randomly assigned item IDs ~\citep{geng2022recommendation};
(10) \textbf{TIGER} also fine-tunes LMs to predict item IDs directly, but these IDs are semantic, meaning they are learned based on the content of the items~\citep{rajput2024recommender}; and
(11) \textbf{IDGenRec} goes further by extending semantic IDs to textual IDs, enriching the IDs with more detailed information~\citep{tan2024idgenrec}.

\paragraph{Implementation Details.}
Unless otherwise specified, we use Gecko \texttt{text-embedding-004}~\citep{lee2024gecko}\footnote{https://cloud.google.com/vertex-ai/generative-ai/docs/model-reference/text-embeddings-api} to collect LLM embeddings for retrieval, and we use \texttt{gemini-1.5-flash}\footnote{https://deepmind.google/technologies/gemini/flash/} for LLM-based ranking.
All API calls were completed as of September 1, 2024.

\begin{table*}[t]
    \begin{minipage}{\textwidth}
	\centering
	\small
	\resizebox{\textwidth}{!}{
		\begin{tabular}{llccccccccccccccc}
            \toprule
           \multirow{2}{*}{\textbf{Category}} & \multirow{2}{*}{\textbf{Method / Model}} & \multirow{2}{*}{\textbf{Train}} & \multicolumn{4}{c}{\textbf{Beauty}} & \multicolumn{4}{c}{\textbf{Toys and Games}} & \multicolumn{4}{c}{\textbf{Sports and Outdoors}}\\
            \cmidrule(lr){4-7} \cmidrule(lr){8-11} \cmidrule(lr){12-15}  & & & H@5 & N@5 & H@10 & N@10 & H@5 & N@5 & H@10 & N@10 & H@5 & N@5 & H@10 & N@10 \\
            \midrule
            \multirow{11}{*}{\texttt{Baseline}} & \texttt{KNN} & \cmark & 0.004 &	0.003 &	0.007 &	0.004 &	0.004 &	0.003 &	0.007 &	0.004 &	0.001 &	0.001 &	0.002 &	0.001  \\
             & \texttt{Caser}~\citep{tang2018personalized} & \cmark & 0.021 & 0.013 & 0.035 & 0.018 & 0.017 & 0.011 & 0.027 & 0.014 & 0.012 & 0.007 & 0.019 & 0.010  \\
             & \texttt{HGN}~\citep{ma2019hierarchical} & \cmark & 0.033 & 0.021 & 0.051 & 0.027 & 0.032 & 0.022 & 0.050 & 0.028 & 0.019 & 0.012 & 0.031 & 0.016 \\
             & \texttt{GRU4Rec}~\citep{hidasi2015session} & \cmark & 0.016 & 0.010 & 0.028 & 0.014 & 0.010 & 0.006 & 0.018 & 0.008 & 0.013 & 0.009 & 0.020 & 0.011 \\
             & \texttt{BERT4Rec}~\citep{sun2019bert4rec} & \cmark & 0.020 & 0.012 & 0.035 & 0.017 & 0.012 & 0.007 & 0.020 & 0.010 & 0.012 & 0.008 & 0.019 & 0.010 \\
             & \texttt{FDSA}~\citep{zhang2019feature} & \cmark & 0.027 & 0.016 & 0.041 & 0.021 & 0.023 & 0.014 & 0.038 & 0.019 & 0.018 & 0.012 & 0.029 & 0.016\\
             & \texttt{SASRec}~\citep{kang2018self} & \cmark & 0.039 & 0.025 & 0.061 & 0.032 & 0.046 & 0.031 & 0.068 & 0.037 & 0.023 & 0.015 & 0.035 & 0.019 \\
             & \texttt{$\bf{S^{3}}$-Rec}~\citep{zhou2020s3} & \cmark & 0.039 & 0.024 &  0.065 & 0.033 & 0.044 & 0.029 & 0.070 & 0.038 & 0.025 & 0.016 &  0.039 & 0.020 \\
             & \texttt{P5}~\citep{zhou2020s3} & \cmark & 0.016 &	0.011 & 0.025 &	0.014 &	0.007 &	0.005 &	0.012 &	0.007 &	0.006 &	0.004 &	0.010 &	0.005 \\
             & \texttt{TIGER}~\citep{geng2022recommendation} & \cmark & 0.045 &	0.032 &	0.065 &	0.038 &	0.052 &	0.037 &	0.071 &	0.043 &	0.026 &	0.018 &	0.040 &	0.023 \\
             & \texttt{IDGenRec}~\citep{tan2024idgenrec} & \cmark & 0.062 & \underline{0.049} &	0.081 &	0.054 &	0.066 &	0.048 &	0.087 &	0.055 &	\bf 0.043 &	\bf 0.033 &	\bf 0.057 &	\bf 0.037 \\
            \midrule
            \texttt{STAR-Retrieval} & \texttt{-} & \xmark & \underline{0.068} &	0.048 &	\underline{0.098} &	\underline{0.057} &	\underline{0.086} &	\underline{0.061} &	\underline{0.118} &	\underline{0.071} &	0.038 &	0.026 &	0.054 & 0.031 \\
            \midrule
            \multirow{3}{*}{\texttt{STAR-Ranking}} & \texttt{point-wise} & \xmark & \underline{0.068} &	0.047 &	0.096 &	0.056 &	\underline{0.086} &	\underline{0.061} &	0.117 &	\underline{0.071} & 0.037 &	0.026 &	0.054 &	0.031 \\
             & \texttt{pair-wise} & \xmark & \bf 0.072 & \bf 0.051 & \bf 0.101 &	\bf 0.060 &	\bf 0.090 &	\bf 0.064 &	\bf 0.120 &	\bf 0.073 &	\underline{0.040} &	\underline{0.028} &	\underline{0.056} &	\underline{0.034}  \\
             & \texttt{list-wise} & \xmark & 0.065 &	0.047 &	0.090 &	0.055 &	0.083 &	0.060 &	0.111 &	0.069 &	0.036 &	0.026 &	0.052 &	0.031 \\
            \bottomrule
        \end{tabular}
	}
	\end{minipage}
	\caption{
	\textbf{Performance (Hits@K, NDCG@K)} comparison among supervised models, and \texttt{STAR retrieval} \& \texttt{ranking} pipeline.
	The first group in the table represents supervised models;
	The second group shows the retrieval pipeline with parameters set to an exponential decay rate of $\lambda=0.7$, history length of $l=3$, and a weight factor of $a=0.5$;
	The third group consists of ranking pipeline which use \texttt{gemini-1.5-flash}.
    The best model for each dataset is shown in \textbf{bold}, and the second best is \underline{underlined}.
    }
	\label{tab:main-table}
\end{table*}

\section{Experimental Results}
In this section, we present a performance analysis of our retrieval (\secref{ssec:performance-retrieval}) and ranking pipeline (\secref{ssec:performance-reranking}).
\subsection{Retrieval Pipeline}
\label{ssec:performance-retrieval}

\begin{figure}[t]
  \begin{center}
    \resizebox{0.55\linewidth}{!}{
      \includegraphics{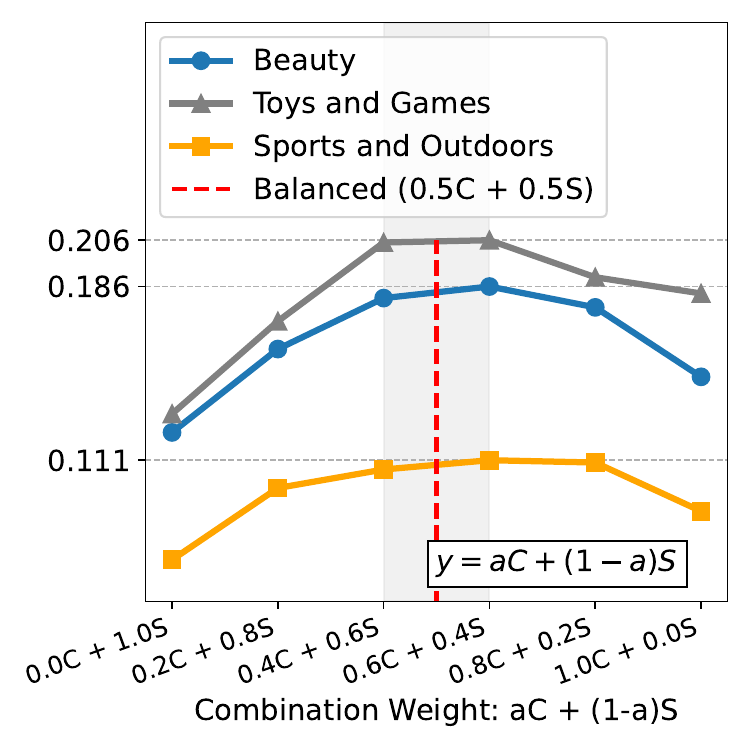}
    }
    
    \resizebox{\linewidth}{!}{
      \includegraphics{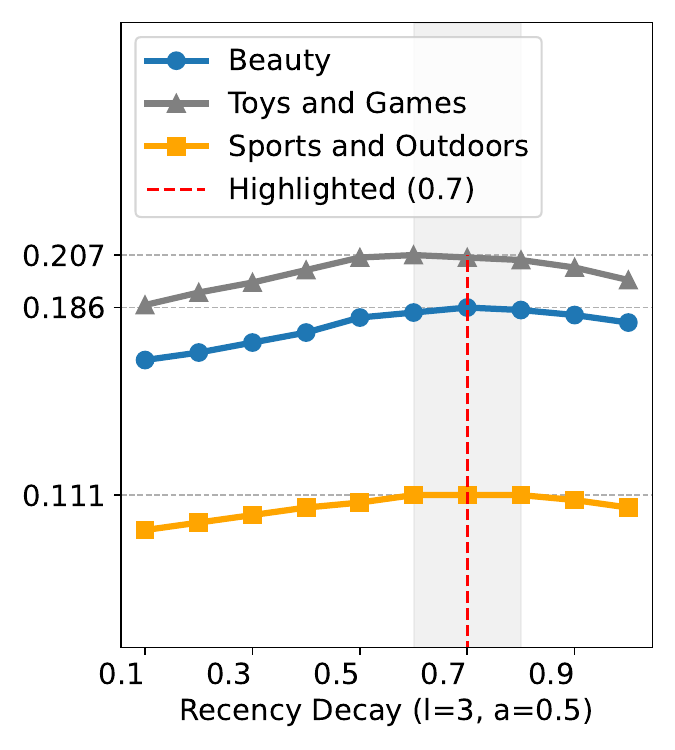}
      \includegraphics{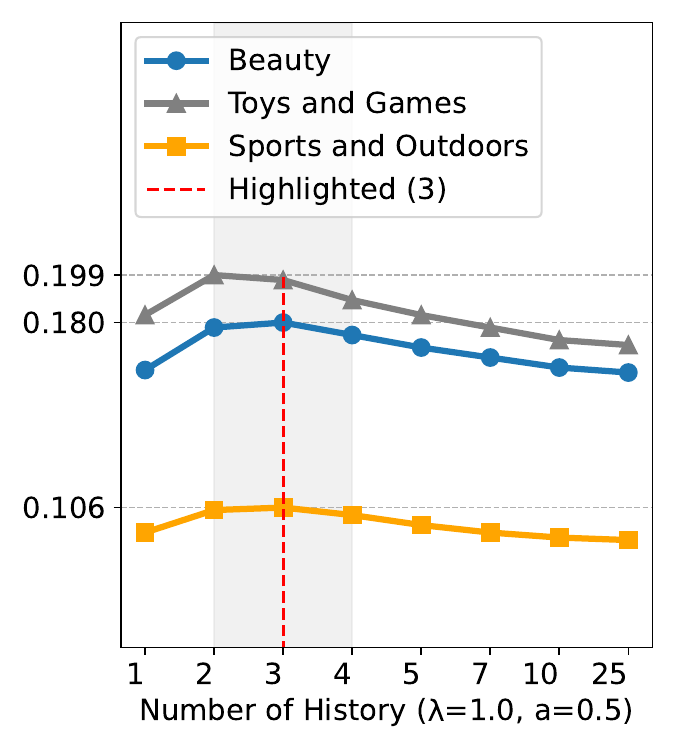}
    }
  \end{center}
  \vspace{-0.4cm}
  \caption{\textbf{Retrieval performance (Hits@50)} with different weighting factor $a$ between $R_{\text{S}}$ and $R_{\text{C}}$ (\textbf{top}), recency factor $\lambda$ (\textbf{bottom-left}), and number of history $l$ (\textbf{bottom-right}). The shaded regions show the best range. $a=0.5$, $\lambda=0.7$, and $l=3$ show the best.}
  \label{fig:factor-analysis}
\vspace{-0.4cm}
\end{figure}

The second group of Table~\ref{tab:main-table} presents the performance of our retrieval framework. \texttt{STAR-Retrieval} alone achieves the best or second-best results compared to all baselines and fine-tuned methods.
There is a significant improvement across all metrics for \textit{Toys and Games}, ranging from +26.50\% to +35.3\%.
In \textit{Beauty}, all metrics besides NDCG@5 (-1.2\%) are improved from a range of +6.1\% to +20.0\%.
In \textit{Sports and Outdoors}, the results are second best to IDGenRec, trailing from a range of -19.6\% to -5.57\%.
Furthermore, we explore the following questions in greater detail, with additional details provided in Appendix~\ref{appendix:retrieval-performance}:

\paragraph{Impact of semantic and collaborative information.}
We assess the impact of semantic ($R_{\text{S}}$) and collaborative ($R_{\text{C}}$) information by varying their weighting factor $a$.
As shown in the top panel of Figure~\ref{fig:factor-analysis}, the optimal performance occurs between $a=0.5$ and $0.6$, with $a=0.5$ chosen for simplicity. The far left and right sides of that figure show that results significantly degrade when using only one component ($a=0.0$ or $a=1.0$), confirming that combining both enhances retrieval effectiveness.

\paragraph{Impact of user history length $l$ and recency factor $\lambda$.}
We analyze the effect of user history length ($l$) and recency factor ($\lambda$) on retrieval performance.
As shown in the right panel of Figure~\ref{fig:factor-analysis}, performance improves with more history items up to $l=3$ but declines beyond that.
The middle panel indicates that a recency factor of $\lambda=0.7$ effectively prioritizes recent interactions, outperforming no recency adjustment ($\lambda=1$).
However, an excessively small $\lambda$ overly discounts past items, resembling the effect of using only one history item.

\paragraph{Impact of user rating.}
\begin{table}[t]
    \begin{minipage}{0.48\textwidth}
    \centering
    \small
    \resizebox{\textwidth}{!}{
    	\begin{tabular}{lccccccc}
    	    \toprule
            \multirow{2}{*}{\textbf{Rating}} & \multicolumn{2}{c}{\textbf{Beauty}} & \multicolumn{2}{c}{\textbf{Toys \& Games}} & \multicolumn{2}{c}{\textbf{Sports \& Outdoors}} \\
            \cmidrule(lr){2-3} \cmidrule(lr){4-5} \cmidrule(lr){6-7} & H@10 & N@10 & H@10 & N@10 & H@10 & N@10 \\
            \midrule
            w/ rating & 0.095 &	0.056 & 0.115 &	0.069 & 0.052 &	0.030 \\
            w/o rating & \textbf{0.098} & \textbf{0.057} & \textbf{0.118} & \textbf{0.071} & \textbf{0.054} &	\textbf{0.031} \\
            \bottomrule
        \end{tabular}
    }
    \end{minipage}
      \caption{\textbf{Retrieval (Hits@10, NDCG@10)} comparing w/ and w/o incorporating user rating.}
    \label{tab:rating}
\end{table}
In Equation ~\ref{eq:1}, we propose using the user ratings to help score items. 
To assess their impact, we compare results using actual ratings ($r$) versus a uniform rating ($r=1$ for all items).
Surprisingly, ignoring the rating information consistently produced better results (See Table~\ref{tab:rating}).
This likely due to a task mismatch—our focus is on predicting interactions, not user ratings.
Consequently, in our main evaluation and the rest of the analysis, we disregard ratings (by setting $r=1$ for all items), which is also practical since most real-world interactions lack explicit ratings.

\paragraph{Scoring rule analysis.}
\begin{table}[t]
    \begin{minipage}{0.48\textwidth}
	\centering
	\small
	\resizebox{\textwidth}{!}{
		\begin{tabular}{lcccccc}
		    \toprule
          \multirow{2}{*}{\textbf{Scoring method}} & \multicolumn{2}{c}{\textbf{Beauty}} & \multicolumn{2}{c}{\textbf{Toys \& Games}} & \multicolumn{2}{c}{\textbf{Sports \& Outdoors}} \\
            \cmidrule(lr){2-3} \cmidrule(lr){4-5} \cmidrule(lr){6-7} & H@10 & N@10 & H@10 & N@10 & H@10 & N@10 \\
            \midrule
            Average Pooling & 0.060	& 0.033 & 0.080	& 0.043 & 0.033 &	0.017 \\
            \texttt{STAR} ($S$=$1.0$, $C$=$0.0$) & 0.072 & 0.042 & 0.095	& 0.055 & 0.039 &	0.022 \\
            \texttt{STAR} ($S$=$0.5$, $C$=$0.5$) & \bf 0.098	& \bf 0.057 & \bf 0.118	& \bf 0.071 & \bf 0.054 &	\bf 0.031  \\
            \bottomrule
        \end{tabular}
	}
	\end{minipage}
	  \caption{\textbf{Retrieval performance (Hits@10, NDCG@10)} comparison between \texttt{STAR-retrieval} pipeline and average embedding pooling. 
	  $S$ representing Semantic Information and $C$ representing Collaborative Information weightings.}
	\label{tab:avg-pooling}
	\vspace{-0.3cm}
\end{table}
Prior studies~\citep{harte2023leveraging, liang2024taxonomy} use LLM embeddings of items in a user sequence ($S_u$) to retrieve candidates.
These methods generate a ``user embedding'' by averaging item embeddings, then identify the most similar candidate based on embedding similarity.
To evaluate our scoring rule, we compare it against this pooling approach.
As shown in Table~\ref{tab:avg-pooling}, our method outperforms pooling even without collaborative information ($S=1.0$, $C=0.0$).
Incorporating collaborative information ($S=0.5$, $C=0.5$) further improves performance, highlighting the advantage of our approach.

\begin{table*}[t!]
    \begin{minipage}{\textwidth}
	\centering
	\small
	\resizebox{\textwidth}{!}{
		\begin{tabular}{lcccccccccccccccc}
            \toprule
           \multirow{2}{*}{\textbf{Prompt Style}} & \multirow{2}{*}{\textbf{Window Size}} & \multirow{2}{*}{\textbf{Stride}} & \multicolumn{4}{c}{\textbf{Beauty}} & \multicolumn{4}{c}{\textbf{Toys and Games}} & \multicolumn{4}{c}{\textbf{Sports and Outdoors}}\\
            \cmidrule(lr){4-7} \cmidrule(lr){8-11} \cmidrule(lr){12-15}  & & & H@5 & N@5 & H@10 & N@10 & H@5 & N@5 & H@10 & N@10 & H@5 & N@5 & H@10 & N@10 \\
            \midrule
            None (STAR-Retrieval) & - & - & 0.0684 & 0.0480 & 0.0977 & 0.0574 & 0.0857 & 0.0606 & 0.1176 & 0.0709 & 0.0379 & 0.0262 & 0.0542 & 0.0314 \\
            \midrule
            Selection & - & - & 0.0691 & 0.0484 & 0.0958 & 0.0570 & 0.0841 & 0.0613 & 0.1109 & 0.0699 & 0.0376 & 0.0269 & 0.0520 & 0.0316 \\
            \midrule
            Point-wise & 1 & 1 & 0.0685 & 0.0472 & 0.0956 & 0.0558 & 0.0855 & 0.0611 & 0.1170 & 0.0713 & 0.0370 & 0.0257 & 0.0539 & 0.0312 \\
            \midrule
            Pair-wise & 2 & 1 & \underline{0.0716} & \textbf{0.0506} & \textbf{0.1008} & \textbf{0.0600} & \textbf{0.0899} & \textbf{0.0639} & 0.1196 & \textbf{0.0734} & \underline{0.0401} & \textbf{0.0283} & \textbf{0.0564} & \textbf{0.0335} \\
            \midrule
            \multirow{4}{*}{List-wise} & 4 & 2 & \textbf{0.0724} & \underline{0.0502} & \underline{0.1002} & \underline{0.0592} & \underline{0.0894} & \underline{0.0634} & 0.1195 & \underline{0.0732} & \textbf{0.0406} & \underline{0.0282} & \underline{0.0559} & \underline{0.0331} \\
             & 8 & 4 & 0.0688 & 0.0484 & 0.0988 & 0.0581 & 0.0874 & 0.0625 & \textbf{0.1202} & 0.0731 & 0.0388 & 0.0276 & 0.0556 & 0.0330 \\
             & 10 & 5 & 0.0676 & 0.0480 & 0.0981 & 0.0578 & 0.0853 & 0.0616 & \underline{0.1201} & 0.0728 & 0.0379 & 0.0270 & 0.0558 & 0.0327 \\
             & 20 & - & 0.0653 & 0.0471 & 0.0903 & 0.0551 & 0.0829 & 0.0603 & 0.1113 & 0.0694 & 0.0364 & 0.0262 & 0.0518 & 0.0311 \\
            \bottomrule
            
        \end{tabular}
	}
	\end{minipage}
    \caption{\textbf{Ranking performance (Hits@K, NDCG@K)} by window size and stride. Here we use 20 candidates from the retrieval stage. The best prompt for each dataset is shown in \textbf{bold}, and the second best is \underline{underlined}.}
    \vspace{-0.2cm}
	\label{tab:window-size}
\end{table*}

\subsection{Ranking Pipeline}
\label{ssec:performance-reranking}
The third group of Table~\ref{tab:main-table} highlights the performance of the ranking framework, which improves upon the retrieval stage results. 
Pair-wise ranking improves all metrics over \texttt{STAR-Retrieval} performance by +1.7\% to +7.9\%. This further improves the results over other baselines for \textit{Beauty} and \textit{Toys and Games}, while closing the gap on IDGenRec in \textit{Sports and Outdoors}. Point-wise and list-wise methods struggle to achieve similar improvements. 
We examine the following questions in more detail, with additional information available in Appendix~\ref{appendix:ranking-performance}.

\paragraph{Effectiveness of ranking and impact of window size and stride.}
Previous approaches use a \textit{selection prompt}, instructing the LLM to choose the top-$k$ items from a candidate set~\citep{wang2023zero, hou2024large, wang-etal-2024-recmind}.
In contrast, our method uses a \textit{ranking} prompt, which explicitly instructs the LLM to rank all items within the available context window. 
We assess the effectiveness of the ranking approach in comparison to the selection approach and a point-wise prompt. 
As shown in Table~\ref{tab:window-size}, results show that ranking with a small window size (\textit{e.g.,} pair-wise or list-wise with a window of 4) consistently outperforms selection and pair-wise approaches.
These findings align with prior work in document retrieval, where list-wise ranking with large windows or reasoning prompts is challenging for LLMs due to the need for task-specific knowledge. In contrast, pair-wise ranking is relatively easier and can yield better performance, even for smaller LMs~\citep{qin-etal-2024-large}.

\paragraph{Impact of candidate count ($k$) and history length ($l$).}
\begin{figure}[t]
\begin{minipage}{0.23\textwidth}
  \centerline{\includegraphics[width=3.8cm]{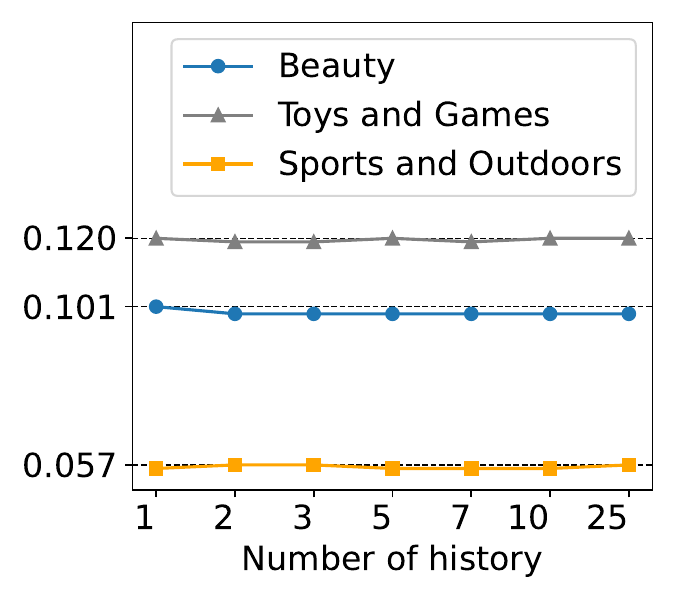}}
  \vspace{-0.1cm}
  \centerline{\small (a) Number of history $l$}
\end{minipage}
\begin{minipage}{0.23\textwidth}
  \centerline{\includegraphics[width=3.8cm]{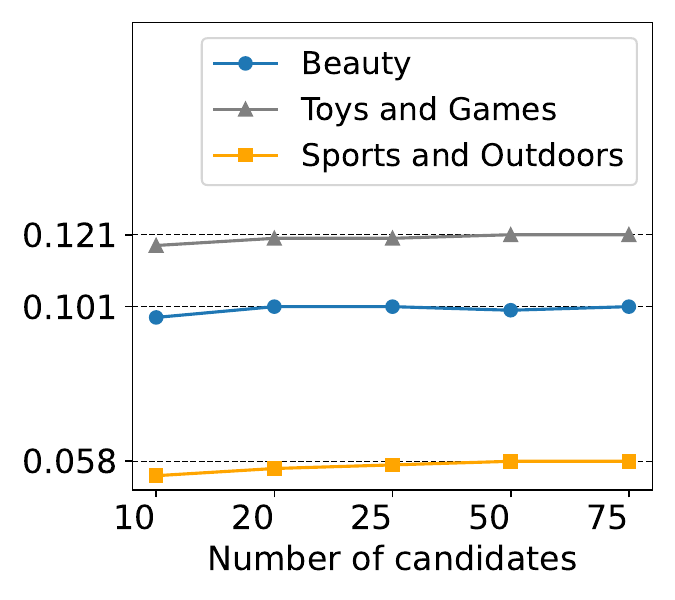}}
  \vspace{-0.1cm}
  \centerline{\small (b) Number of candidates $k$}
\end{minipage}
\vspace{-0.2cm}
\caption{\textbf{Pair-wise ranking performance (Hits@10)} trend by different number of history $l$ and number of candidates $k$}
\label{fig:ranking-history-candidate}
\end{figure}
As shown in Figure~\ref{fig:ranking-history-candidate}, varying the number of historical items ($l$) has little effect on performance.
In contrast, increasing the number of candidate items ($k$) slightly improves accuracy by increasing the chances of including the correct item.
However, the performance gains are minimal, and given the high computational cost of processing more candidates, the trade-off may not be worthwhile.

\paragraph{Impact of additional item information in prompts.}
\begin{table}[t]
    \begin{minipage}{0.48\textwidth}
	\centering
	\small
	\resizebox{\textwidth}{!}{
		\begin{tabular}{lccccccc}
		    \toprule
            \multirow{2}{*}{\textbf{Item prompt}} & \multicolumn{2}{c}{\textbf{Beauty}} & \multicolumn{2}{c}{\textbf{Toys \& Games}} & \multicolumn{2}{c}{\textbf{Sports \& Outdoors}} \\
            \cmidrule(lr){2-3} \cmidrule(lr){4-5} \cmidrule(lr){6-7} & H@10 & N@10 & H@10 & N@10 & H@10 & N@10 \\
            \midrule
            Metadata & 0.1000 &	0.0567 & 0.1193 & 0.0690 & 0.0544 &	0.0315 \\
            + popularity & 0.0998 & 0.0564 & 0.1174 & 0.0701 & 0.0549 &	0.0316 \\
            + co-occurrence & \bf 0.1008 &	\bf 0.0600 & 0.1196 &	0.0734 & \bf 0.0564 &	\bf 0.0335   \\
            + popularity, co-occurrence & 0.0999 & 	0.0599 & \textbf{0.1203}	& \textbf{0.0736} & 0.0550 & 0.0322 \\
            \bottomrule
        \end{tabular}
	}
	\end{minipage}
	  \caption{\textbf{Pair-wise ranking (Hits@10 \& NDCG@10)} by varying information in the item prompt.}
	\label{tab:prompt-information}
	\vspace{-0.4cm}

\end{table}
We analyze the effect of adding extra item information—specifically, popularity and co-occurrence data—alongside item metadata in prompts.
As shown in Table~\ref{tab:prompt-information}, incorporating co-occurrence data improves NDCG@10 by +0.2\% to +3.4\% compared to using metadata alone.
In contrast, adding popularity information does not enhance performance and sometimes even reduces it.
This suggests that popularity bias does not help LLMs make better ranking decisions, consistent with prior research showing its ineffectiveness in recommendation tasks~\citep{abdollahpouri2020multi, abdollahpouri2019unfairness, zhu2021popularity}.

\paragraph{Impact of candidate order.}

We investigate whether the order of the retrieval candidate items affects the ranking performance of our recommendation system. 
To assess this, we conducted an experiment comparing the pairwise ranking outcomes between two sets of top 20 candidates:
(1) Random order: Candidates were randomly shuffled; (2) Retrieval Order: Candidates were ordered based on their scores from the retrieval pipeline.
As shown in Table~\ref{tab:rank-order}, ordering the candidates according to their retrieval pipeline scores significantly improves ranking performance compared to a random arrangement. 
Comparing rows 2 and 4 show that our pair-wise ranking can improve results of a randomly shuffled list. 
However, the results are even better when the candidate list is ranked by the retrieval score (row 3). 
More analysis needs to be done to determine if $O(n\log n)$ or $O(n^2)$ comparisons could better rank a randomly shuffled candidate list compared to a sliding window approach, although this would come at an even higher computation cost.

\begin{table}[t]
    \begin{minipage}{0.48\textwidth}
	\centering
	\small
	\resizebox{\textwidth}{!}{
		\begin{tabular}{cccccccccc}
		    \toprule
            \multirow{2}{*}{\textbf{Shuffle}} &
            \multirow{2}{*}{\textbf{LLM}} &
            \multicolumn{2}{c}{\textbf{Beauty}} & \multicolumn{2}{c}{\textbf{Toys \& Games}} & \multicolumn{2}{c}{\textbf{Sports \& Outdoors}} \\
            \cmidrule(lr){3-4} \cmidrule(lr){5-6} \cmidrule(lr){7-8} \textbf{Candidates} & \textbf{Ranking} & H@10 & N@10 & H@10 & N@10 & H@10 & N@10 \\
            \midrule
            \xmark & \xmark & 0.0977 & 0.0574 & 0.1176 & 0.0709 & 0.0542 & 0.0314 \\
            \cmark & \xmark & 0.0687 & 0.0312 & 0.0779 & 0.0349 & 0.0371 & 0.0169 \\
            \xmark & \cmark & 0.1008 & 0.0600 & 0.1196 & 0.0734 & 0.0564 & 0.0335 \\
            \cmark & \cmark & 0.0793 & 0.0485 & 0.0949 & 0.0596 & 0.0452 & 0.0275 \\
            \bottomrule
        \end{tabular}
	}
	\end{minipage}
	  \caption{\textbf{Ranking (Hits@10 \& NDCG@10)}
	  comparison with random shuffling of the retrieved items.}
	\label{tab:rank-order}
	\vspace{-0.4cm}
\end{table}
\section{Conclusion}
In this paper, we introduced a \textbf{S}imple \textbf{T}raining-free \textbf{A}pproach for \textbf{R}ecommendation (\textbf{STAR}) that uses the power of large language models (LLMs) to create a generalist framework applicable across multiple recommendation domains.
Our method comprises two key stages: a retrieval phase and a ranking phase. In the retrieval stage, we combine semantic embeddings from LLMs with collaborative user information to effectively select candidate items. In the ranking stage, we apply LLMs to enhance next-item prediction and refine the recommendations.
Experimental results on a large-scale Amazon review dataset demonstrate that our retrieval method alone outperforms most supervised models.
By employing LLMs in the ranking stage, we achieve further improvements.
Importantly, our study highlights that incorporating collaborative information is critical in both stages to maximize performance.
Our findings reveal that LLMs can effectively function as generalists in recommendation tasks without requiring any domain-specific fine-tuning. 
This opens up exciting possibilities for developing versatile and efficient recommendation systems that are readily adaptable across diverse domains.

\section{Limitations}
The \textbf{STAR} framework presents an effective alternative to traditional supervised models, showcasing the potential of LLMs in recommendation systems without the need for extensive training or custom architectures. 
However, several limitations remain, which also indicate directions for future improvement:

\paragraph{Importance of item modality and enriched item meta-data.}
The \textbf{STAR} framework's ability to capture semantic relationships between items relies significantly on the presence of rich item text meta-data.
Without such meta-data and with only user-item interaction data available, the framework’s semantic relationship component will be less effective.
To maximize the use of semantic relationships between items, future work should explore incorporating additional modalities, such as visual or audio data, to generate more comprehensive semantic representations of items, fully utilizing all the available information.

\paragraph{Improving Retrieval Simplicity and Scalability.} Although our work demonstrates the effectiveness of a general training-free framework, the current method requires different choices for parameters. In future work, we will explore ways to either reduce the number of parameters choices or select values more easily. In our current implementation, we compute the full set of item-item comparisons for both the semantic and collaborative information. This computation is infeasible if the item set is too large. In future work, we will run experiments to measure how effective approximate nearest neighbor methods are at reducing computation and maintaining retrieval quality.


\paragraph{Beyond LLM ranking.}
The importance of our work highlights that high quality results can be achieved without additional fine-tuning. However, in the current method, our \textbf{STAR} ranking pipeline utilizes costly LLM calls that would result in high latency. This may be a suitable solution to use in offline scenarios, but would be prohibitive to serve large-scale and real-time user traffic. Future work needs to explore how we can improve efficiency, such as using a mix of pair-wise and list-wise ranking. Our work shows a promising first step to creating high quality, training-free, and general recommendation systems.

\subsubsection*{Acknowledgments}
We would like to thank Yupeng Hou (UCSD), Jinhyuk Lee (Google DeepMind), Jay Pujara (USC ISI), Jiaxi Tang (Google DeepMind) and Wang-Cheng Kang (Google DeepMind) for their valuable discussions and contributions to this work.


\bibliography{custom}
\clearpage
\appendix

\definecolor{systemcolor}{RGB}{230, 230, 255} 
\definecolor{usercolor}{RGB}{255, 240, 230}   
\definecolor{assistantcolor}{RGB}{235, 255, 235} 

\newmdenv[
    backgroundcolor=systemcolor, 
    linecolor=black, 
    linewidth=0.5pt, 
    skipabove=1pt, 
    skipbelow=1pt, 
    innertopmargin=3pt, 
    innerbottommargin=3pt
]{systembox}

\newmdenv[
    backgroundcolor=usercolor, 
    linecolor=black, 
    linewidth=0.5pt, 
    skipabove=1pt, 
    skipbelow=1pt, 
    innertopmargin=3pt, 
    innerbottommargin=3pt
]{userbox}

\newmdenv[
    backgroundcolor=assistantcolor, 
    linecolor=black, 
    linewidth=0.5pt, 
    skipabove=1pt, 
    skipbelow=1pt, 
    innertopmargin=3pt, 
    innerbottommargin=3pt
]{assistantbox}

\section{Appendix}
\subsection{Retrieval Performance Analysis}
\label{appendix:retrieval-performance}
\begin{figure}[h]
  \begin{center}
    \includegraphics[width=\linewidth]{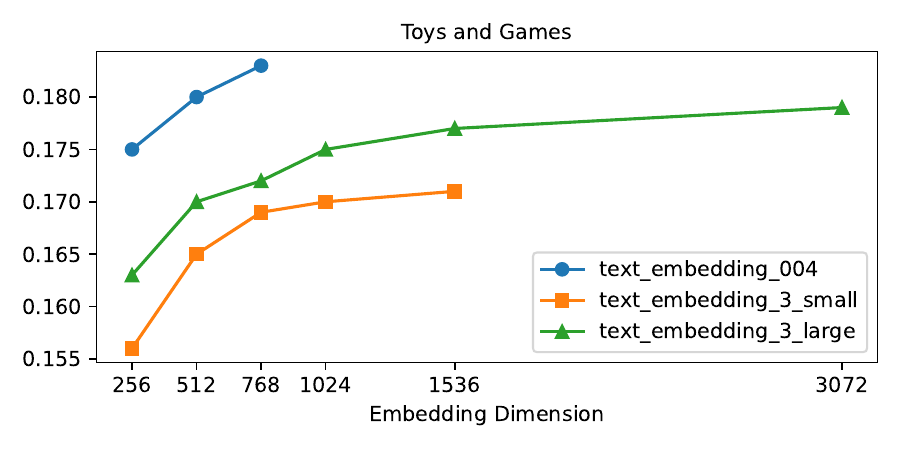}
  \end{center}
    \vspace{-0.3cm}
  \caption{\textbf{Retrieval performance (Hits@50)} comparison by embedding APIs for Toys and Games. The models text-embedding-004, text-embedding-3-small, and text-embedding-3-large each have a maximum dimension of 768, 1536, and 3072, respectively.}
  \vspace{-0.3cm}
  \label{fig:embedding}
\end{figure}

\paragraph{LLM embedding performance comparison on retrieval.}
We test different LLM embedding APIs with different embedding size to understand the impact of LLM embeddings capturing item semantic similarity .
Figure~\ref{fig:embedding} illustrates the performance differences between the Gecko \texttt{text-embedding-004}~\cite{lee2024gecko}, OpenAI \texttt{text-embedding-3-small}, and \texttt{text-embedding-3-large} models\footnote{https://platform.openai.com/docs/guides/embeddings/}. Overall, higher-dimensional embeddings and larger models performed better, indicating that enhanced semantic representation capabilities lead to improved semantic relationship capture.

\subsection{Ranking Performance Analysis}
\label{appendix:ranking-performance}

\paragraph{LLM performance comparison on ranking.}
To evaluate how model capabilities impact ranking performance, we compare four models: \texttt{gemini-1.5-flash}, \texttt{gemini-1.5-pro}, \texttt{gpt-4o-mini}, and \texttt{gpt-4o}.
In Table~\ref{tab:llm-model}, we observe the larger models (\texttt{gemini-1.5-pro} and \texttt{gpt-4o}) tend to outperform their smaller counterparts, although the performance differences are minimal. This finding aligns with results from other studies~\citep{qin-etal-2024-large}, which suggest that increased model capability has limited influence on pairwise ranking tasks. 
Despite being more computationally expensive, pairwise ranking methods tend to be more robust than alternative approaches.

\begin{table}[h]
    \begin{minipage}{0.48\textwidth}
	\raggedright
	\small
	\resizebox{\textwidth}{!}{
		\begin{tabular}{lccccccc}
		    \toprule
            \multirow{2}{*}{\textbf{Model}} & \multicolumn{2}{c}{\textbf{Beauty}} & \multicolumn{2}{c}{\textbf{Toys \& Games}} & \multicolumn{2}{c}{\textbf{Sports \& Outdoors}} \\
            \cmidrule(lr){2-3} \cmidrule(lr){4-5} \cmidrule(lr){6-7} & H@10 & N@10 & H@10 & N@10 & H@10 & N@10 \\
            \midrule
            \texttt{gemini-1.5-flash} & \bf 0.101 & 0.060 & 0.120 & 0.073 & \bf 0.056 & 0.034\\
            \texttt{gemini-1.5-pro} & 0.100 &  0.060 & 0.120 & \bf 0.075 & 0.056 & \bf 0.034\\
            \texttt{gpt-4o-mini} & 0.100 & 0.058 & 0.120 & 0.072 & 0.056 & 0.033\\
            \texttt{gpt-4o} & 0.100 & \textbf{0.060} & \textbf{0.121} & 0.074 & 0.056 & 0.033 \\
            \bottomrule
        \end{tabular}
	}
	\end{minipage}
	  \caption{\textbf{Pair-wise ranking performance (Hits@10 \& NDCG@10)} comparison by different models.}
	\label{tab:llm-model}
\end{table}

\onecolumn
\subsection{Item Encoding Prompt for Retrieval Example}
\label{app:item-prompt}
Below is an example of an item prompt for encoding with an LLM embedding API. This example includes fields for \textit{title}, \textit{description}, \textit{category}, \textit{brand}, \textit{sales ranking}, and \textit{price}. We omit metadata fields like \textit{Item ID} and \textit{URL}, as those fields contain strings that can contain spurious lexical similarity (e.g., IDs: ``000012'', ``000013' or URLs: ``https://abc.com/uxrl'', ``https://abc.com/uxrb'') and can reduce the uniformity of the embedding space and make it difficult to distinguish between semantically different items 

\vspace{0.5cm}
\begin{userbox}
\small
\begin{verbatim}
description: 
    LENGTH: 70cm / 27.55 inches
    Color: Mix Color
    EST. SHIPPING WT.: 310g
    Material: Synthetic High Temp Fiber
    Cap Construction: Capless
    Cap Size: Average
    1. The size is adjustable and no pins or tape should be required. It should fit most people.
       Adjust the hooks inside the cap to suit your head.
    2. Please be aware that colors might look slightly different in person due to
       camera quality and monitor settings. Stock photos are taken in natural light with no flash.
    3. Please ask all questions prior to purchasing. I will replace defective items. 
       Indicate the problem before returning.
       A 30-day return/exchange policy is provided as a satisfaction guarantee.

title: 63cm Long Zipper Beige+Pink Wavy Cosplay Hair Wig Rw157
salesRank: {'Beauty': 2236}
categories: Beauty > Hair Care > Styling Products > Hair Extensions & Wigs > Wigs
price: 11.83
brand: Generic
\end{verbatim}
\end{userbox}
\clearpage

\subsection{Ranking Prompt Example}
\label{app:ranking-prompt}
Below is an example of a single pass in a list-wise ranking pipeline with a window size of 4 and a stride of 2 ($w=4$ and $d=2$) assuming there are 3 history items ($l=3$).

\vspace{0.5cm}

\begin{systembox}
\small \textbf{System:} You are an intelligent assistant that can rank items based on the user's preference.
\end{systembox}

\begin{userbox}
\small \textbf{User:} User 1656 has purchased the following items in this order:

\begin{verbatim}
{
    "Item ID": 1069,
    "title": "SHANY Professional 13-Piece Cosmetic Brush Set with Pouch, 
    Set of 12 Brushes and 1 Pouch, Red",
    "salesRank_Beauty": 248,
    "categories": [
        ["Beauty", "Tools & Accessories", "Makeup Brushes & Tools", "Brushes & Applicators"]
    ],
    "price": 12.95,
    "brand": "SHANY Cosmetics"
},
{
    "Item ID": 2424,
    "title": "SHANY Eyeshadow Palette, Bold and Bright Collection, Vivid, 120 Color",
    "salesRank_Beauty": 1612,
    "categories": [
        ["Beauty", "Makeup", "Eyes", "Eye Shadow"]
    ],
    "price": 16.99,
    "brand": "SHANY Cosmetics"
},
{
    "Item ID": 2856,
    "title": "SHANY Studio Quality Natural Cosmetic Brush Set with Leather Pouch, 24 Count",
    "salesRank_Beauty": 937,
    "categories": [
        ["Beauty", "Tools & Accessories", "Bags & Cases", "Cosmetic Bags"]
    ],
    "price": 26.99,
    "brand": "SHANY Cosmetics"
}
\end{verbatim}
I will provide you with 4 items, each indicated by number identifier []. Analyze the user's purchase history to identify preferences and purchase patterns. Then, rank the candidate items based on their alignment with the user's preferences and other contextual factors.
\end{userbox}

\begin{assistantbox}
\small \textbf{Assistant:} Okay, please provide the items.
\end{assistantbox}

\begin{userbox}
\small \textbf{User:} [1]
\begin{verbatim}
{
    "title": "SHANY Cosmetics Intense Eyes Palette 72 Color Eyeshadow Palette, 17 Ounce",
    "salesRank_Beauty": 181358,
    "categories": [
        ["Beauty", "Makeup", "Makeup Sets"]
    ],
    "price": 26.4,
    "brand": "SHANY Cosmetics",
    "Number of users who bought both this item and Item ID 1069": 18,
    "Number of users who bought both this item and Item ID 2424": 0,
    "Number of users who bought both this item and Item ID 2856": 16
}
\end{verbatim}
\end{userbox}

\begin{assistantbox}
\small \textbf{Assistant:} Received item [1].
\end{assistantbox}

\begin{userbox}
\small \textbf{User:} [2]
\begin{verbatim}
{
    "title": "SHANY Cosmetics Carry All Train Case with Makeup and Reusable Aluminum Case, Cameo",
    "salesRank_Beauty": 2439,
    "categories": [
        ["Beauty", "Makeup", "Makeup Sets"]
    ],
    "price": 39.99,
    "brand": "SHANY Cosmetics",
    "Number of users who bought both this item and Item ID 1069": 27,
    "Number of users who bought both this item and Item ID 2424": 1,
    "Number of users who bought both this item and Item ID 2856": 29
}
\end{verbatim}
\end{userbox}

\begin{assistantbox}
\small \textbf{Assistant:} Received item [2].
\end{assistantbox}

\begin{userbox}
\small \textbf{User:} [3]
\begin{verbatim}
{
    "title": "SHANY COSMETICS The Masterpiece 7 Layers All-in-One Makeup Set",
    "salesRank_Beauty": 2699,
    "categories": [
        ["Beauty", "Makeup", "Makeup Sets"]
    ],
    "price": 41.89,
    "brand": "SHANY Cosmetics",
    "Number of users who bought both this item and Item ID 1069": 23,
    "Number of users who bought both this item and Item ID 2424": 2,
    "Number of users who bought both this item and Item ID 2856": 25
}
\end{verbatim}
\end{userbox}

\begin{assistantbox}
\small \textbf{Assistant:} Received item [3].
\end{assistantbox}



\begin{userbox}
\small \textbf{User:} [4]
\begin{verbatim}
{
    "title": "SHANY Silver Aluminum Makeup Case, 4 Pounds",
    "salesRank_Beauty": 16605,
    "categories": [
        ["Beauty", "Tools & Accessories", "Bags & Cases", "Train Cases"]
    ],
    "price": 59.95,
    "brand": "SHANY Cosmetics",
    "Number of users who bought both this item and Item ID 1069": 32,
    "Number of users who bought both this item and Item ID 2424": 1,
    "Number of users who bought both this item and Item ID 2856": 40
}
\end{verbatim}
\end{userbox}

\begin{assistantbox}
\small \textbf{Assistant:} Received item [4].
\end{assistantbox}

\begin{userbox}
\small \textbf{User:} Analyze the user's purchase history to identify user preferences and purchase patterns.
Then, rank the 4 items above based on their alignment with the user's preferences and other contextual factors.
All the items should be included and listed using identifiers, in descending order of the user's preference.
The most preferred recommendation item should be listed first.
The output format should be [] > [], where each [] is an identifier, e.g., [1] > [2].
Only respond with the ranking results, do not say any word or explain.
Output in the following JSON format:
\begin{verbatim}
{
    "rank": "[] > [] .. > []"
}
\end{verbatim}
\end{userbox}

\end{document}